\documentclass[pdflatex,sn-nature]{sn-jnl}
\usepackage[T1]{fontenc}
\usepackage[utf8]{inputenc}
\usepackage{graphicx,amsmath,amssymb,textcomp,microtype,xurl}
\usepackage{float,fancyhdr}
\usepackage{enumitem}
\usepackage{xcolor,soul}
\makeatletter
\renewcommand\normalsize{\@setfontsize\normalsize{11pt}{14pt}\abovedisplayskip=8pt\belowdisplayskip=8pt}
\makeatother
\hypersetup{hidelinks}
\fancypagestyle{plain}{\fancyhf{}\fancyfoot[C]{\thepage}}
\begin{document}
\normalsize

% SOURCE_PARAGRAPH 1
\begin{center}{\LARGE\bfseries Planar energetics do not determine twin persistence in nanocrystalline aluminium\par}\vspace{10pt}\end{center}

% SOURCE_PARAGRAPH 3
\begin{center}Yue Wu\textsuperscript{1,2}, Yuan Tian\textsuperscript{2}, Haoyu Wan\textsuperscript{1}, Deng Pan\textsuperscript{1,*}, Mingwei Chen\textsuperscript{2,*}\end{center}\vspace{-4pt}

% SOURCE_PARAGRAPH AUTHOR_AFFILIATION_1
{\centering\fontsize{10}{12}\selectfont \textsuperscript{1}Materials Genome Institute, Shanghai University, Shanghai 200444, China\par}

% SOURCE_PARAGRAPH AUTHOR_AFFILIATION_2
{\centering\fontsize{10}{12}\selectfont \textsuperscript{2}Department of Materials Science and Engineering, Southern University of Science and Technology, Shenzhen 518055, China\par}

% SOURCE_PARAGRAPH AUTHOR_CORRESPONDENCE
{\centering\fontsize{9.5}{12}\selectfont \textsuperscript{*}Correspondence: \url{DPan_MGI@shu.edu.cn} (Deng Pan); \url{chenmw@sustech.edu.cn} (Mingwei Chen)\par}

% SOURCE_PARAGRAPH AUTHOR_EMAILS
{\centering\fontsize{9.5}{12}\selectfont Author e-mails: \url{repeatsongyue@gmail.com} (Yue Wu); \url{tiany3@sustech.edu.cn} (Yuan Tian); \url{wanhaoyu0715@gmail.com} (Haoyu Wan)\par}\vspace{8pt}

% SOURCE_PARAGRAPH 7,9
\noindent\textbf{Abstract}\par\smallskip
Planar energetics are the standard descriptors of twinnability in face-centred cubic metals and the usual validation targets for interatomic potentials. Here we show that they do not determine whether deformation twins persist in nanocrystalline aluminium. Two machine-learned potentials reproducing the same zero-load slip curves and closely spaced layered energy surfaces predict, from a common separated-fault precursor under identical loading, retained twin coverage of 64\% and 14\%. The divergence traces to forces within distorted partial cores acting along the trailing translation that would remove an outer fault. Planar energetics do set the margin: an outer fault more than halves the incremental barrier for middle-plane shear, yet twin completion and fault removal remain separated by only 10--21 mJ m\textsuperscript{-2}. Whether an assembled twin persists is decided instead by the orientation of local traction relative to the full-slip bisector and by the arrival order of partial-slip fronts from the grain-boundary network.

% SOURCE_PARAGRAPH 11
\section*{Introduction}

% SOURCE_PARAGRAPH 12
The idea that stacking-fault energy controls deformation twinning in nanocrystalline metals was sharpened by Van Swygenhoven, Derlet and Frøseth, who argued that the relevant quantity is not the intrinsic stacking-fault energy alone but its ratio to the unstable stacking-fault energy, which sets whether a grain-boundary-emitted partial can separate from its trailing partner before the fault is removed \textsuperscript{1}. That argument reframed twinning as a competition between two partial-dislocation processes acting on the same fault, and it established planar energetics --- intrinsic and unstable stacking-fault energies, and their combinations --- as the standard descriptors of twinnability. Aluminium became the canonical test case precisely because this competition is acute there: a high intrinsic stacking-fault energy (\textasciitilde{}122 mJ m\textsuperscript{-2}) is paired with a modest unstable-slip peak (\textasciitilde{}159 mJ m\textsuperscript{-2}), leaving a shallow well of only \textasciitilde{}36 mJ m\textsuperscript{-2} that stabilizes a fault against trailing-partial recovery \textsuperscript{1,2}. Twinnability measures built on these quantities, and the shear-direction analyses that followed, inherit the same logic: whether a twin forms is read off an energy landscape computed for a perfect crystal under uniform load \textsuperscript{2,8--10}.

% SOURCE_PARAGRAPH 13
Two features of the deforming polycrystal are not represented in that logic. The first is that the transformation advances through localized partial-dislocation cores within distorted boundary neighbourhoods, where the force that moves the defect is a property of the core, not of the planar energy surface \textsuperscript{3,4}. Errors of even a few tens of mJ m\textsuperscript{-2} in slip barriers, or subtle biases in core forces, can invert the balance between twin completion and fault removal \textsuperscript{1}, and gigapascal normal stresses near grain boundaries shift fault energies by amounts comparable to the shallow Al well \textsuperscript{9,10,12,13}. The second is that the fault being acted on is not prepared in isolation. In a nanometre-sized grain, partials traverse short distances between boundaries~\textsuperscript{3--6} and encounter neighbouring defects, so the stacking a given plane presents to an incoming partial records slip that arrived from elsewhere; grain orientation and loading influence which slip system operates \textsuperscript{8--10}, while boundary deformation changes the sources and propagation paths that supply it.

% SOURCE_PARAGRAPH 14
The 1--3--2 route identified by Wang and colleagues makes the second feature explicit \textsuperscript{7}. Across three successive \{111\} planes P1, P2 and P3, faults form first on the outer planes P1 and P3, after which shear on the intervening P2 completes the twin; the adjacent-plane 1--2--3 route instead builds the embryo by consecutive slip. From a single stacking fault, slip on the adjacent plane forms a thin embryo, whereas slip one plane farther away creates a separated-fault pair whose middle-plane shear has a reduced energy increment \textsuperscript{7}. The energetic advantage of the 1--3--2 route therefore depends on first creating an additional outer fault, which links the ease of twin completion to the cost of precursor preparation. This is where planar energetics alone become insufficient as a predictor: the same three planes support two routes, the choice between them depends on the spatial order in which faults arrive, and that order is not contained in any single-plane energy curve.

% SOURCE_PARAGRAPH 15
Here we test whether planar energetics determine twin persistence. We use Al-QP, a MACE-based potential optimized via constrained quadratic programming on strain-coupled fault, twin and interface responses, benchmarked against five existing models on published DFT data and paired with first-principles defect calculations. We first show that two machine-learned potentials reproducing the same zero-load slip curves and closely spaced layered energy surfaces predict sharply different retained twin coverage in the polycrystal, and trace the difference to forces within moving partial cores. We then resolve how a local 1--3--2 sequence emerges from partial-slip fronts arriving asynchronously from different neighbourhoods, and relate local traction to the evolving stacking: near the full-slip bisector, the loading that creates an outer fault is already aligned with the translation that would remove it. Finally, we show that an outer fault more than halves the incremental barrier for middle-plane shear while storing energy in a competing removal channel, leaving twin completion and fault removal separated by only 10--21 mJ m\textsuperscript{-2}. Planar energetics set that margin but do not decide the outcome; moving-core forces and front arrival do.

% SOURCE_PARAGRAPH 21
\section*{Results}

% SOURCE_PARAGRAPH 22
\subsection*{Matched planar energetics, divergent twin persistence}

% SOURCE_PARAGRAPH 23
We take planar energetics to mean the intrinsic stacking-fault energy, the unstable stacking-fault energy and the twin energy --- the quantities that enter twinnability measures and against which interatomic potentials are conventionally validated \textsuperscript{1,2}. Our quadratic-program-optimized potential (Al-QP) and the Li--Mishin Al moment-tensor potential (Al-MTP) were developed with different objectives. Al-QP was trained on strain-coupled fault, twin and interface responses, including ordinary-slip, successive-twinning and incoherent-boundary configurations under load. Al-MTP was developed to investigate normal-stress effects on stacking-fault energy \textsuperscript{12}. Their training therefore emphasizes different parts of configuration space, and any agreement between them on planar energetics is not built in by construction.

% SOURCE_PARAGRAPH 24
Both reproduce the zero-load single-plane slip energetics of the Andric--Yin--Curtin DFT reference closely \textsuperscript{13} (Fig. 1a). Their intrinsic fault energies are 121.05 and 122.42 mJ m\textsuperscript{-2}, against 122.49 mJ m\textsuperscript{-2} from DFT; their leading-slip maxima are 150.61 and 154.76 mJ m\textsuperscript{-2}, against 158.80 mJ m\textsuperscript{-2}. On the layered 48-atom reference carrying the local deformation of the polycrystal event analysed below, the two potentials also give closely spaced energy profiles along both competing channels (Fig. 1b). For P2 shear with the outer fault present, the incremental maxima are 26.84 and 29.27 mJ m\textsuperscript{-2}; with the outer fault absent they are 72.73 and 71.23 mJ m\textsuperscript{-2}. At q = 0.3, the P3-minus-P2 energy separations are 2.27 and 4.49 mJ m\textsuperscript{-2} for Al-QP and Al-MTP, respectively (differing by 2.21 mJ m\textsuperscript{-2}, and by 3.24 mJ m\textsuperscript{-2} at q = 0.5). Along both zero-load single-plane slip and the selected competing layered-crystal paths, the two descriptions are nearly interchangeable.

% SOURCE_PARAGRAPH 25
They are not interchangeable in the polycrystal. We continued both potentials from a common separated-fault precursor, the 19.50 ps configuration of a 1--3--2 event in a 6.0 nm polycrystal at 200 K, with shared thermal velocities and shared imposed loading --- and followed the central stacking for 5 ps. Across eight additional paired thermal-velocity realizations, Al-QP reached completed central stacking in all eight and retained it at every terminal configuration; Al-MTP reached completion in five of eight and retained it in three of seven available terminal configurations (Fig. 1d). Mean resolved twin coverage over the surrounding region, measured on 37 fixed precursor windows and 100 time samples, was approximately 64\% with Al-QP and 14\% with Al-MTP (Fig. 1e). Both potentials initiate middle-plane shear and can assemble the twin, yet Al-MTP can stall during assembly or exhibit only intermittent completed stacking. The divergence therefore concerns both the completion of assembly and the persistence of the assembled twin.

% SOURCE_PARAGRAPH 26
The P3 displacement identifies the competing route that consumes the outer fault. Under Al-MTP and Mishin-1999 EAM, continued trailing slip removes the upper fault and recovers parent stacking; under Mendelev-2008 EAM the twin persists throughout the 5 ps continuation, and under Al-QP it persists until late-stage stacking changes after approximately 4.4 ps (Fig. 1f; Supplementary Fig. 13a). These histories identify trailing-partial recovery on the outer plane as a competing process that can consume the fault supporting twin assembly. Because the two potentials reproduce the selected layered P2 and P3 energy profiles almost identically, this divergence points beyond planar energetics to the local force that drives a partial dislocation through its distorted core.

% SOURCE_PARAGRAPH 27
That force differs between the models. We extracted finite clusters around a P3 core during continued fault-removing slip and compared the slip-force imbalance \(D=(\langle F_{\mathrm{upper}}\rangle-\langle F_{\mathrm{lower}}\rangle)\cdot s\) with DFT references at cluster radii of 8, 10 and 12 Å (Fig. 1g). Al-MTP gives a systematically stronger forward drive along the trailing translation than DFT, by +0.04 to +0.08 eV Å\textsuperscript{-1} across the three radii; Al-QP lies within \(-\)0.03 to +0.02 eV Å\textsuperscript{-1} of the same references. At the P2 core during twin completion, by contrast, the two potentials agree to approximately 0.022 eV Å\textsuperscript{-1} and both lie in the range spanned by the DFT cluster sizes (Fig. 1h). The models thus differ specifically along the fault-removal channel, and the sign of the difference matches the direction of the observed divergence: the potential with the stronger trailing drive retains fewer twins.

% SOURCE_PARAGRAPH 28
These comparisons establish the paper's central observation. Planar energetics --- the quantities that twinnability measures are built from and that potentials are conventionally validated against --- are matched between Al-QP and Al-MTP to within a few mJ m\textsuperscript{-2}. Twin persistence in the polycrystal nevertheless differs by a factor of four. The difference is carried by forces inside moving partial cores, along the competing translation that removes the outer fault. Matching zero-load slip curves or layered-crystal energy surfaces does not, by itself, determine what a deforming polycrystal does.

% SOURCE_PARAGRAPH 30
\subsection*{Directional traction and front arrival determine which plane transforms}

% SOURCE_PARAGRAPH 31
Tensile molecular dynamics covered nanocrystalline Al with nominal grain diameters of 4.7--7.5 nm at 100--500 K. Trajectories contain both adjacent-plane twinning and separated-fault assembly; six positions across three polycrystals follow the local 1--3--2 order (Supplementary Table 5). We resolve a representative event in a 6.0 nm polycrystal at 200 K between 6 and 11\% axial strain (Fig. 2a,b). A fault on P1 forms at 12.40--12.80 ps. A second fault on P3, already active elsewhere in the grain, advances and overlaps the P1 fault at the centre between 18.15 and 18.75 ps while P2 retains parent stacking (Fig. 2c,d,f,g,i). On P2, a lower neighbouring region transforms between 17.45 and 18.15 ps; the front subsequently completes the twin at the central position between 19.80 and 20.50 ps (Fig. 2e,h,j,k). The completed region has a median parent misorientation of 59.45\(^{\circ}\) and a median deviation of 1.65\(^{\circ}\) from the corresponding \(\Sigma\)3 relation.

% SOURCE_PARAGRAPH 32
The local order is therefore not generated by a single sequential source (Supplementary Fig. 12). The outer-plane front arrives from elsewhere before the middle-plane front, so fault overlap prepares the precursor geometry and the arriving P2 front completes it. Which plane transforms first is set by front arrival, not by the magnitude of inter-plane shear. The same sequence occurs at 300 and 400 K (Supplementary Table 5). This is the kinetic half of what determines persistence: a separated-fault precursor exists only if an outer-plane front reaches the position before the middle-plane front does, and its lifetime is the interval between the two arrivals.

% SOURCE_PARAGRAPH 33
Local stress biases which crystallographic family is favoured on a given plane. Projecting the in-plane traction onto the six signed a/6\(\langle\)112\(\rangle\) Shockley translations permitted by the current layer stacking defines a stacking-shear compatibility: for plane normal n and configurational stress \(\sigma\), \(\tau_a=s_a\cdot t_l\) with \(t_l=(I-n\otimes n)\cdot\sigma\cdot n\). The three translations that advance the selected reversed sequence are separated by 120\(^{\circ}\); their opposites traverse non-twinning configurations. In the representative 1--3--2 event and a second site, P3 experiences \textasciitilde{}1 GPa shear along a twin-compatible direction, directly driving the observed outer-plane slip. In adjacent twin thickening, by contrast, the strongest shear on P3 lies along the non-twinning opposite family while slip proceeds on P2 along the compatible direction (Fig. 3a--c). Across sampled environments, P3 loading in 1--3--2 sites shifts toward the twin-compatible family, whereas both adjacent-plane cases prefer the opposite family (Fig. 3d).

% SOURCE_PARAGRAPH 34
Resolving the traction against the Shockley directions also exposes the mechanical origin of the subsequent competition. Ahead of the arriving P3 front, the mean in-plane traction --- accompanied by 1.5--1.6 GPa normal tension --- lies approximately 1.3\(^{\circ}\) from the full a/2\(\langle\)110\(\rangle\) slip direction that bisects the leading translation D0 and the 60\(^{\circ}\)-rotated D1 (Fig. 3a). D0 and D1 therefore receive nearly identical resolved shears of 1.06 and 1.09 GPa. Before P3 slips, D0 is the twin-compatible leading translation that creates the outer fault; once that fault forms and reverses the stacking, D1 becomes the trailing route that erases it. The loading geometry that promotes fault formation is already aligned with the translation that can remove it. This near-degeneracy is the directional half of what determines persistence: the same local stress state that creates the precursor also supplies the drive for its removal, and only the arrival of the middle-plane front can pre-empt that removal.

% SOURCE_PARAGRAPH 35
Together, the two conditions explain the divergence in Section 1. Planar energetics set the energy margin between completion and removal but do not decide which side of it the system lands on. Traction orientation decides whether an outer fault forms and whether the loading already favours its removal; front arrival decides whether the middle-plane partial reaches the position before the trailing partial does. Neither is encoded in a zero-load slip curve.

% SOURCE_PARAGRAPH 37
\subsection*{A narrow energetic margin leaves the outcome to traction and arrival}

% SOURCE_PARAGRAPH 38
An outer fault changes both the energy required for middle-plane shear and the energy stored in the precursor. In the constrained layered reference carrying the representative event's local deformation, P2 shear from a single fault (100 \(\rightarrow\) 110) encounters an incremental energy maximum of 72.73 mJ m\textsuperscript{-2} with Al-QP and 71.23 mJ m\textsuperscript{-2} with Al-MTP. This adjacent-plane path ends with a near-zero net energy change. With P3 already faulted (101 \(\rightarrow\) 111), the maximum falls by more than half to 26.84 and 29.27 mJ m\textsuperscript{-2} before the path descends to a lower-energy twin state (Fig. 4a). The reduction is preserved across the 300 and 400 K event-derived deformation references (Supplementary Fig. 14). An outer fault therefore substantially lowers the incremental barrier for middle-plane shear, in agreement with the first-principles comparison of Wang et al. \textsuperscript{7}.

% SOURCE_PARAGRAPH 39
The reduction comes at a cost. Relative to the same single-fault state (100), the DFT separated-fault state (101) stores an extra 125.85 mJ m\textsuperscript{-2} (Fig. 4b,c). The outer fault thus facilitates twin completion while supplying stored energy that trailing slip on P3 can also release --- two competing consequences of the same precursor geometry. P2 converts the fault pair into a twin; P3 returns it to a single fault. Both transformations are energy-lowering from the common precursor, and both are accessible to the same local stress state identified in Section 2.

% SOURCE_PARAGRAPH 40
Two-dimensional energy surfaces resolve both routes from the common 101 precursor (Fig. 4d,e). Along P2 leading slip (q\textsubscript{2}) and P3 trailing slip (q\textsubscript{3}), the model surfaces rise and then descend into separate valleys toward the completed twin and upper-fault-removed states. An elevated interior ridge penalizes simultaneous uniform shear on both planes. At each matched partial displacement examined (q = 0.3, 0.5 and 1.0), DFT places the P2 twin-completing route below P3 fault removal (Fig. 4f). At 30\% and 50\% of a partial translation, middle-plane shear is lower in energy by 10.45 and 20.64 mJ m\textsuperscript{-2}, respectively. The completed twin lies 139.35 mJ m\textsuperscript{-2} below the precursor, gaining 13.50 mJ m\textsuperscript{-2} more than upper-fault removal.

% SOURCE_PARAGRAPH 41
These separations are small. For a constant illustrative shear difference of 0.1 GPa over a partial displacement q b\textsubscript{p}, with b\textsubscript{p} = 1.66 Å, the work difference per unit slipped area is 16.6q mJ m\textsuperscript{-2}. This gives 4.98 and 8.30 mJ m\textsuperscript{-2} at q = 0.3 and 0.5, respectively, and 16.6 mJ m\textsuperscript{-2} over a full partial translation. At q = 0.3, the illustrative work exceeds the models' own channel separations (2.27 and 4.49 mJ m\textsuperscript{-2}); over a full translation, it exceeds the 13.50 mJ m\textsuperscript{-2} DFT endpoint separation (Fig. 4f). This places local mechanical driving on the same energy scale as the competing layered-crystal profiles: planar energetics set an energetic preference, but the narrow margin allows local loading to compete with it. The same comparison explains why the 64\%/14\% coverage difference in Section 1 does not contradict the layered-crystal energetics: both potentials reproduce a narrow margin, while their differing force responses within moving cores provide a local driving bias. Together with traction orientation and front arrival in Section 2, these force responses connect the energetic comparison to the contrasting assembly and persistence histories (Fig. 5a--d).

% SOURCE_PARAGRAPH 42
Al-QP and Al-MTP give closely spaced profiles along both channels, with smaller channel separations than DFT (Fig. 4a,d,e; Supplementary Note 2). At q = 0.5 the P2 energies exceed the DFT value by 28.48 and 30.85 mJ m\textsuperscript{-2}. The models therefore reproduce the qualitative structure of the competing channels --- two energy-lowering routes separated by a narrow margin --- while compressing that margin. Compressing the margin does not change the conclusion; it strengthens it, because a narrower margin makes the outcome more sensitive to the non-energetic conditions identified in Section 2.

% SOURCE_PARAGRAPH 43
Taken together, Sections 1--3 separate three conditions that planar energetics conflate (Fig. 6). The energy landscape sets a margin between twin completion and fault removal, and an outer fault makes that margin narrow. Traction orientation sets whether the loading favours formation and removal simultaneously, as it does near the full-slip bisector. Front arrival sets whether the middle-plane partial reaches the position before the trailing partial does. Planar energetics determine the first; they do not determine the second or third. Because the first leaves only a narrow margin, the second and third decide the outcome.

% SOURCE_PARAGRAPH 45
\section*{Discussion}

% SOURCE_PARAGRAPH 46
The criterion and its three conditions. Our results separate three conditions that planar-energetics frameworks conflate. The energy landscape sets a margin between twin completion and fault removal, and an outer fault makes that margin narrow: an existing P3 fault more than halves the incremental barrier for middle-plane shear, yet DFT places twin completion below fault removal by only 10--21 mJ m\textsuperscript{-2}, and the separated-fault precursor stores 125.85 mJ m\textsuperscript{-2} that trailing slip can also release. Traction orientation sets whether the loading favours formation and removal simultaneously; near the full-slip bisector the leading and trailing Shockley translations receive nearly equal resolved shear, so the geometry that creates an outer fault is already aligned with the translation that erases it. Front arrival sets whether the middle-plane partial reaches the position before the trailing partial does, and therefore whether a separated-fault precursor exists long enough to be completed. Planar energetics determine the first condition; they do not determine the second or third. Because the first leaves only a narrow margin, the second and third decide the outcome. We do not claim that planar energetics are irrelevant: they set the energetic margin within which the outcome is decided, and they favour twin completion over fault removal. The claim is that they are insufficient as predictors of persistence, because the margin is narrow enough that traction orientation, front arrival and core forces decide which side of it the system lands on.

% SOURCE_PARAGRAPH 47
Implications for potential validation. The comparison between Al-QP and Al-MTP carries a methodological conclusion that extends beyond aluminium twinning. The two potentials reproduce the same zero-load slip curves and closely spaced layered-crystal energy surfaces, yet predict retained twin coverage of 64\% and 14\% from a common precursor under identical loading. The difference is carried by forces inside moving partial cores along the trailing translation that removes the outer fault: Al-MTP supplies a stronger forward drive than DFT by +0.04 to +0.08 eV Å\textsuperscript{-1} across cluster radii, whereas the two potentials agree to approximately 0.022 eV Å\textsuperscript{-1} at the P2 core during twin completion. Matching planar energetics is therefore not sufficient evidence that two potentials will agree on defect evolution in a deforming polycrystal. Because errors of a few tens of mJ m\textsuperscript{-2} in slip barriers or subtle biases in core forces can invert the balance between completion and removal \textsuperscript{1}, potentials intended for competing partial-slip processes require validation on both strained multi-fault energetics and local forces in distorted defect cores. The 64\%/14\% divergence is not a failure of either potential on its own validation set; it is a demonstration that the conventional validation set does not constrain the quantity that decides persistence.

% SOURCE_PARAGRAPH 48
Rate and transferability. Separating source initiation, front propagation and precursor evolution clarifies how loading rate enters atomistic interpretations of twinning in nanocrystalline Al \textsuperscript{1,3--6}. Kinetically, the high strain rates of molecular dynamics influence the stress sustained near grain boundaries, the activation of boundary sources, and the time window over which an arriving outer-plane front overlaps an existing fault before middle-plane completion or trailing-partial erasure. At fixed local deformation and boundary constraints, however, the conditional energy landscape and the geometric stacking-shear compatibility relations contain no explicit dependence on nominal strain rate. The frequency with which separated-fault precursors assemble and survive therefore depends on rate-sensitive front kinetics, whereas the reduced barrier and the narrow DFT preference provide an intrinsic energetic basis for the evolution of an assembled precursor. This separation is what makes the criterion transferable: it identifies which ingredients are rate-dependent (front arrival, source activation) and which are not (traction orientation, the energy margin), so the same framework can be applied at experimental rates by replacing the simulated arrival statistics with measured ones. The framework should also transfer to other high-stacking-fault-energy FCC metals, where the same competition between partials operates and the same near-bisector geometry arises; Al is the most stringent test case because its fault-energy well is shallow, not a special case.

% SOURCE_PARAGRAPH 49
A testable signature. The criterion makes predictions accessible to in situ transmission electron microscopy of nanocrystalline Al. First, 1--3--2 assembly requires an outer fault to persist while a middle-plane front arrives; the predicted window between outer-fault formation and middle-plane completion should widen at lower strain rates, where front-arrival lags grow, and narrow at higher rates. Second, because the traction that creates the outer fault lies near the full-slip bisector, fault formation and removal are mechanically nearly indistinguishable at that orientation; twin initiation should therefore drop sharply for grain orientations loading P3 near this bisector, and the drop should be sharper in metals with a narrower energetic margin. Third, the divergent retained coverage predicted by Al-QP and Al-MTP despite matched planar energetics implies that experimental twin fractions should be compared against simulated retention statistics rather than against slip-curve accuracy when discriminating between interatomic models. None of these predictions requires a new simulation to state; each is a consequence of the three conditions above.

% SOURCE_PARAGRAPH 51
Planar energetics are the standard descriptors of twinnability, but they are not predictors of twin persistence. The energy landscape sets a margin; traction orientation and front arrival decide which side of it the system lands on; and forces within moving partial cores carry the difference between potentials that match on every planar observable. Because the same separation applies wherever competing partial-slip processes operate on a near-degenerate landscape, it defines the validation axis that potentials for such processes need: not only the energies of strained multi-fault configurations, but the forces inside the cores that move them.

% SOURCE_PARAGRAPH 54
\section*{Methods}

% SOURCE_PARAGRAPH 55
\subsection*{Dataset preparation and Al-QP training}

% SOURCE_PARAGRAPH 56
Al-QP combines a broad Al database with ordinary-slip, successive-twinning, strain-coupled fault, HCP/DHCP, elastic and incoherent \(\Sigma\)3 configurations. Development labels were calculated with PAW-PBE in VASP \textsuperscript{23--25}, retaining each data family's numerical settings and energy references. Missing stress labels were masked, virial signs reconciled, and corrected elastic constraints constructed after excluding unreliable labels. The MACE architecture \textsuperscript{17} uses two interaction layers, a 5 Å cutoff and hidden representations 128\(\times\)0e + 128\(\times\)1o; atomic contributions sum to a total energy whose derivatives give forces and stress. After conventional fitting, Al-QP fixed the shared representation and optimized 2,192 parameters in three readout-weight tensors under a constrained quadratic program that combined normalized gradients of grouped physical objectives (broad force and energy, interface force, elasticity, path CVaR, strained force and stress, twin energies at 3\% and 5\%, interface relative energies, and function preservation), with checks on each realized update. Four final runs were compared on a 4\%-strain selection set and broad energy-validation error; the selected run used seed 944 and 160 updates. The 4\% set and historical evaluations form part of model development. Supplementary Note 1 records the development sequence, objective definitions, fallback conditions, hyperparameters, dataset counts and model identities.

% SOURCE_PARAGRAPH 58
\subsection*{Published reference benchmarks}

% SOURCE_PARAGRAPH 59
All six potentials were evaluated without refitting. The Andric--Yin--Curtin reference \textsuperscript{13} contains 39 final structures, 13 at each nominal normal tension; relative energies use the authors' extrapolated E0 values and the unslipped structure in each loading family as reference. The near-fault force RMSE in Supplementary Fig. 11c uses Cartesian force components for atoms within 3 Å of the periodic slip plane, with the same atom mask for every potential. Loaded-boundary comparisons \textsuperscript{20,21} distinguish common geometries from model-relaxed structures and retain the published energy or excess-enthalpy definitions, including external work. For the 41-state loaded-boundary series, loading increments are measured relative to the nominal-scale reference of each microstructure; excluding those three reference states leaves 38 increments. Four general boundaries use the published values in ref. 22. Supplementary Tables 1--3 specify sampling, energy conventions and reference sensitivity. Training-overlap information for the external potentials is incomplete, so the comparisons are described as published-reference evaluations.

% SOURCE_PARAGRAPH 61
\subsection*{Tensile deformation and potential substitutions}

% SOURCE_PARAGRAPH 62
Periodic Voronoi cells contained four or eight grains within an initially approximately 96 Å cube, corresponding to nominal grain diameters of 7.5 and 6.0 nm. Temperature and realization coverage, together with supplementary size and loading-direction comparisons, are given in Supplementary Table 4. LAMMPS \textsuperscript{26} used a 1 fs step, 5 ps NVT equilibration and 15 ps isotropic zero-pressure equilibration, followed by 30 ps of z tension. Transverse pressure was controlled during loading. The legacy protocol applied successive segments with a nominal rate parameter of 0.005 ps\textsuperscript{-1} and reset the reference length between segments; reported engineering strains are calculated from actual cell dimensions. The principal specimen starts at Lz = 95.2208185 Å; its 12, 14 and 4 ps segments reach 15.6884\% strain. Supplementary Table 4 lists the later corrected size/direction series separately.

% SOURCE_PARAGRAPH 63
Potential substitutions used the same 19.50 ps physical configuration and newly assigned thermal velocities, without minimization or additional equilibration. One velocity realization was shared by Al-QP, Al-MTP and both EAM models. Eight further paired realizations compared Al-QP with Al-MTP, generated at 200 K using seeds 2026101301--2026101308, removing centre-of-mass and initial rigid-body rotation. Each branch evolved directly for 5 ps without minimization, unloading or additional equilibration, inheriting the box-growth rate Lz(t\(\prime\)) = 104.719095146173 + 0.5046703380538472t\(\prime\) Å with t\(\prime\) in ps. Pairing fixes initial coordinates, velocities and loading; stresses and subsequent motion are calculated independently by each potential. The source trajectory's instantaneous velocities were unavailable. Terminal-configuration availability and the coverage definition are given in Supplementary Note 5.

% SOURCE_PARAGRAPH 65
\subsection*{Structural analysis and electronic-structure comparisons}

% SOURCE_PARAGRAPH 66
Layer stacking, continuous relative displacement and orientation relative to the current parent jointly identify local twinning. Polyhedral template matching and OVITO \textsuperscript{27,28} support the structural analysis. Three-dimensional views retain physical coordinates; labelled material-coordinate slices and displacement maps remove affine deformation. Opaque spheres show whole-cell structure, with sections exposing internal faults. A clear stacking state required a reference distance below 0.60 Å and at least 0.30 Å smaller than the competing distance; joint assignments required agreement between displacement and stacking measures. Principal maps used 4.5 Å support windows with five to eight atoms per plane, checked at 5 and 6 Å. Unresolved assignments form a separate state. Supplementary Note 4 details the stacking assignments, spatial tracking and the material-coordinate definitions for the principal event.

% SOURCE_PARAGRAPH 67
A 48-atom layered reference carries the local deformation of the 19.50 ps precursor. Two prescribed Shockley translations, q\textsubscript{2} on P2 and q\textsubscript{3} on P3, define an 11\(\times\)11 grid over 0 \(\leq\) q\textsubscript{2}, q\textsubscript{3} \(\leq\) 1. Each model relaxes the interior normal coordinates at fixed cell, in-plane positions and outer-layer clamps; energy differences are divided by the common slip-plane area and referenced to that model's relaxed precursor. DFT independently relaxes selected configurations under the same geometric constraints. The adjacent-plane reference was obtained by restoring parent stacking across P3 while retaining the same 48-atom cell, deformation and outer-layer constraints; its P2 cut was re-referenced to the corresponding relaxed single-fault energy. Supplementary Note 2 gives the cell vectors, slip-plane area, q\textsubscript{2}/q\textsubscript{3} displacement definitions, DFT settings and the coordinate equivalence for the adjacent-plane reference.

% SOURCE_PARAGRAPH 68
Finite clusters extracted from two moving defects compare instantaneous atomic forces at fixed physical coordinates. Six core atoms straddle each active interface; their slip-force imbalance is \(D=(\langle F_{\mathrm{upper}}\rangle-\langle F_{\mathrm{lower}}\rangle)\cdot s\), where s is the relative-slip direction. Figures 1g,h and 5 compare full-system model forces with finite-cluster DFT forces. Clusters of radius 8, 10 and 12 Å were extracted from a P3 core during continued fault-removing slip and a P2 core during twin completion, retaining physical coordinates and fixed six-atom cores. Single-point DFT calculations used PBE, 700 eV \(\Gamma\)-point sampling, 0.05 eV Gaussian smearing, fixed ions and an electronic tolerance of 10\textsuperscript{-7} eV; the 8 and 10 Å calculations met the tolerance, while the 12 Å calculations reached the iteration limit and are marked separately. Supplementary Note 5 reports the radius-dependent values, core-force changes and electronic-convergence measurements.

% SOURCE_PARAGRAPH 70
\subsection*{Local stress and crystallographic compatibility}

% SOURCE_PARAGRAPH 71
Six signed Shockley directions were defined in the current parent basis; the three translations from parent to reversed stacking were identified geometrically before evaluating stress. Resolved shear is \(\tau=s\cdot\sigma_{\mathrm{conf}}\cdot n\), with tensile-positive configurational stress averaged over physical spheres of radius 6 or 8 Å using the native MACE edge-force virial partition; kinetic contributions are excluded. Probe centres follow the physical positions of the tracked material-layer interfaces. Across 37 snapshots, summed atomic and global stresses agree to 1.23\(\times\)10\textsuperscript{-5} GPa. Supplementary Note 6 specifies the virial partition, probe-centre fitting, pre-event intervals, averaging definitions and the treatment of coarse sampling and pre-existing slip. The matched-plane comparison uses the same signed Shockley direction on P2 and P3, chosen as the nearest crystallographic direction to the subsequently observed net slip. Samples within one trajectory describe correlated evolution.

% SOURCE_PARAGRAPH 73
\section*{Data and code availability}

% SOURCE_PARAGRAPH 74
Published references are available through the Materials Cloud archive for ref. 13 (\url{https://doi.org/10.24435/materialscloud:2019.0089/v1}) and the Zenodo archives for refs. 20 and 21 (\url{https://doi.org/10.5281/zenodo.6414955} and \url{https://doi.org/10.5281/zenodo.15674815}). The Zenodo record for this work is available at \url{https://doi.org/10.5281/zenodo.23634744} (version v1). The development identifier QP17 denotes Al-QP in Supplementary Note 1 and the original model records. Numerical and structural source data for Figs. 1--6 and Supplementary Figs. 10--15 accompany this manuscript.

% SOURCE_PARAGRAPH 181
\section*{References}
\begin{enumerate}[label={[\arabic*]},leftmargin=2.5em,itemsep=3pt,parsep=0pt]\small\raggedright

% SOURCE_PARAGRAPH 182
\item Van Swygenhoven, H., Derlet, P. M. \& Frøseth, A. G. Stacking fault energies and slip in nanocrystalline metals. Nature Materials 3, 399--403 (2004). \url{https://doi.org/10.1038/nmat1136}

% SOURCE_PARAGRAPH 183
\item Tadmor, E. B. \& Bernstein, N. A first-principles measure for the twinnability of FCC metals. Journal of the Mechanics and Physics of Solids 52, 2507--2519 (2004). \url{https://doi.org/10.1016/j.jmps.2004.05.002}

% SOURCE_PARAGRAPH 184
\item Yamakov, V., Wolf, D., Phillpot, S. R., Mukherjee, A. K. \& Gleiter, H. Dislocation processes in the deformation of nanocrystalline aluminium by molecular-dynamics simulation. Nature Materials 1, 45--49 (2002). \url{https://doi.org/10.1038/nmat700}

% SOURCE_PARAGRAPH 185
\item Yamakov, V., Wolf, D., Phillpot, S. R. \& Gleiter, H. Deformation twinning in nanocrystalline Al by molecular-dynamics simulation. Acta Materialia 50, 5005--5020 (2002). {\setlength{\fboxsep}{0pt}\url{https://doi.org/10.1016/S1359-6454(02)00318-X}}

% SOURCE_PARAGRAPH 186
\item Chen, M. et al. Deformation twinning in nanocrystalline aluminum. Science 300, 1275--1277 (2003). \url{https://doi.org/10.1126/science.1083727}

% SOURCE_PARAGRAPH 187
\item Liao, X. Z. et al. Deformation mechanism in nanocrystalline Al: Partial dislocation slip. Applied Physics Letters 83, 632--634 (2003). \url{https://doi.org/10.1063/1.1594836}

% SOURCE_PARAGRAPH 188
\item Wang, L. et al. New twinning route in face-centered cubic nanocrystalline metals. Nature Communications 8, 2142 (2017). \url{https://doi.org/10.1038/s41467-017-02393-4}

% SOURCE_PARAGRAPH 189
\item McCabe, R. J., Beyerlein, I. J., Carpenter, J. S. \& Mara, N. A. The critical role of grain orientation and applied stress in nanoscale twinning. Nature Communications 5, 3806 (2014). \url{https://doi.org/10.1038/ncomms4806}

% SOURCE_PARAGRAPH 190
\item Jo, M., Koo, Y. M., Lee, B.-J., Johansson, B., Vitos, L. \& Kwon, S. K. Theory for plasticity of face-centered cubic metals. Proceedings of the National Academy of Sciences USA 111, 6560--6565 (2014). \url{https://doi.org/10.1073/pnas.1400786111}

% SOURCE_PARAGRAPH 191
\item Sun, H., Ding, Z., Sun, H., Li, S., Lavernia, E. J. \& Liu, W. The synergistic effects of energy barriers and shear directions on twinning in face centered cubic metals. Computational Materials Science 177, 109547 (2020). \url{https://doi.org/10.1016/j.commatsci.2020.109547}

% SOURCE_PARAGRAPH 192
\item Ogata, S., Li, J. \& Yip, S. Energy landscape of deformation twinning in bcc and fcc metals. Physical Review B 71, 224102 (2005). \url{https://doi.org/10.1103/PhysRevB.71.224102}

% SOURCE_PARAGRAPH 193
\item Li, Y. \& Mishin, Y. The effect of normal stress on stacking fault energy in face-centered cubic metals. Acta Materialia 312, 122243 (2026). \url{https://doi.org/10.1016/j.actamat.2026.122243}

% SOURCE_PARAGRAPH 194
\item Andric, P., Yin, B. \& Curtin, W. A. Stress-dependence of generalized stacking fault energies. Journal of the Mechanics and Physics of Solids 122, 262--279 (2019). \url{https://doi.org/10.1016/j.jmps.2018.09.007}

% SOURCE_PARAGRAPH 195
\item Smith, J. S. et al. Automated discovery of a robust interatomic potential for aluminum. Nature Communications 12, 1257 (2021). \url{https://doi.org/10.1038/s41467-021-21376-0}

% SOURCE_PARAGRAPH 196
\item Störmer, I. \& Zavadlav, J. Aluminum solidification and nanopolycrystal deformation via a Graph Neural Network Potential and Million-Atom Simulations. arXiv 2603.24360 (2026). \url{https://arxiv.org/abs/2603.24360}

% SOURCE_PARAGRAPH 197
\item Li, Z. et al. Navigating polymorph generation and distilled-potential development via entropy-symmetry landscapes for metal plasticity mechanisms. Nature Communications 17, 5070 (2026). \url{https://doi.org/10.1038/s41467-026-73188-9}

% SOURCE_PARAGRAPH 198
\item Batatia, I., Kovács, D. P., Simm, G. N. C., Ortner, C. \& Csányi, G. MACE: Higher order equivariant message passing neural networks for fast and accurate force fields. Advances in Neural Information Processing Systems 35, 11423--11436 (2022). {\setlength{\fboxsep}{0pt}\url{https://doi.org/10.52202/068431-0830}}

% SOURCE_PARAGRAPH 199
\item Mishin, Y., Farkas, D., Mehl, M. J. \& Papaconstantopoulos, D. A. Interatomic potentials for monoatomic metals from experimental data and ab initio calculations. Physical Review B 59, 3393--3407 (1999). \url{https://doi.org/10.1103/PhysRevB.59.3393}

% SOURCE_PARAGRAPH 200
\item Mendelev, M. I., Kramer, M. J., Becker, C. A. \& Asta, M. Analysis of semi-empirical interatomic potentials appropriate for simulation of crystalline and liquid Al and Cu. Philosophical Magazine 88, 1723--1750 (2008). \url{https://doi.org/10.1080/14786430802206482}

% SOURCE_PARAGRAPH 201
\item Saood, S., Brink, T., Liebscher, C. H. \& Dehm, G. Microstates and defects of incoherent \(\Sigma\)3 [111] twin boundaries in aluminum. Acta Materialia 243, 118499 (2023). \url{https://doi.org/10.1016/j.actamat.2022.118499}

% SOURCE_PARAGRAPH 202
\item Choi, Y. \& Brink, T. Faceting transition in aluminum as a grain boundary phase transition. Physical Review Materials 9, 083607 (2025). \url{https://doi.org/10.1103/2dnf-zdz8}

% SOURCE_PARAGRAPH 203
\item Zheng, H. et al. Grain boundary properties of elemental metals. Acta Materialia 186, 40--49 (2020). \url{https://doi.org/10.1016/j.actamat.2019.12.030}

% SOURCE_PARAGRAPH 204
\item Perdew, J. P., Burke, K. \& Ernzerhof, M. Generalized gradient approximation made simple. Physical Review Letters 77, 3865--3868 (1996). \url{https://doi.org/10.1103/PhysRevLett.77.3865}

% SOURCE_PARAGRAPH 205
\item Blöchl, P. E. Projector augmented-wave method. Physical Review B 50, 17953--17979 (1994). \url{https://doi.org/10.1103/PhysRevB.50.17953}

% SOURCE_PARAGRAPH 206
\item Kresse, G. \& Furthmüller, J. Efficient iterative schemes for ab initio total-energy calculations using a plane-wave basis set. Physical Review B 54, 11169--11186 (1996). \url{https://doi.org/10.1103/PhysRevB.54.11169}

% SOURCE_PARAGRAPH 207
\item Thompson, A. P. et al. LAMMPS: A flexible simulation tool for particle-based materials modeling at the atomic, meso, and continuum scales. Computer Physics Communications 271, 108171 (2022). \url{https://doi.org/10.1016/j.cpc.2021.108171}

% SOURCE_PARAGRAPH 208
\item Larsen, P. M., Schmidt, S. \& Schiøtz, J. Robust structural identification via polyhedral template matching. Modelling and Simulation in Materials Science and Engineering 24, 055007 (2016). \url{https://doi.org/10.1088/0965-0393/24/5/055007}

% SOURCE_PARAGRAPH 209
\item Stukowski, A. Visualization and analysis of atomistic simulation data with OVITO---the Open Visualization Tool. Modelling and Simulation in Materials Science and Engineering 18, 015012 (2010). \url{https://doi.org/10.1088/0965-0393/18/1/015012}

\end{enumerate}

\clearpage

% NEW_FIGURE Fig1
\begin{figure}[p]\centering
\includegraphics[width=\textwidth,height=0.79\textheight,keepaspectratio]{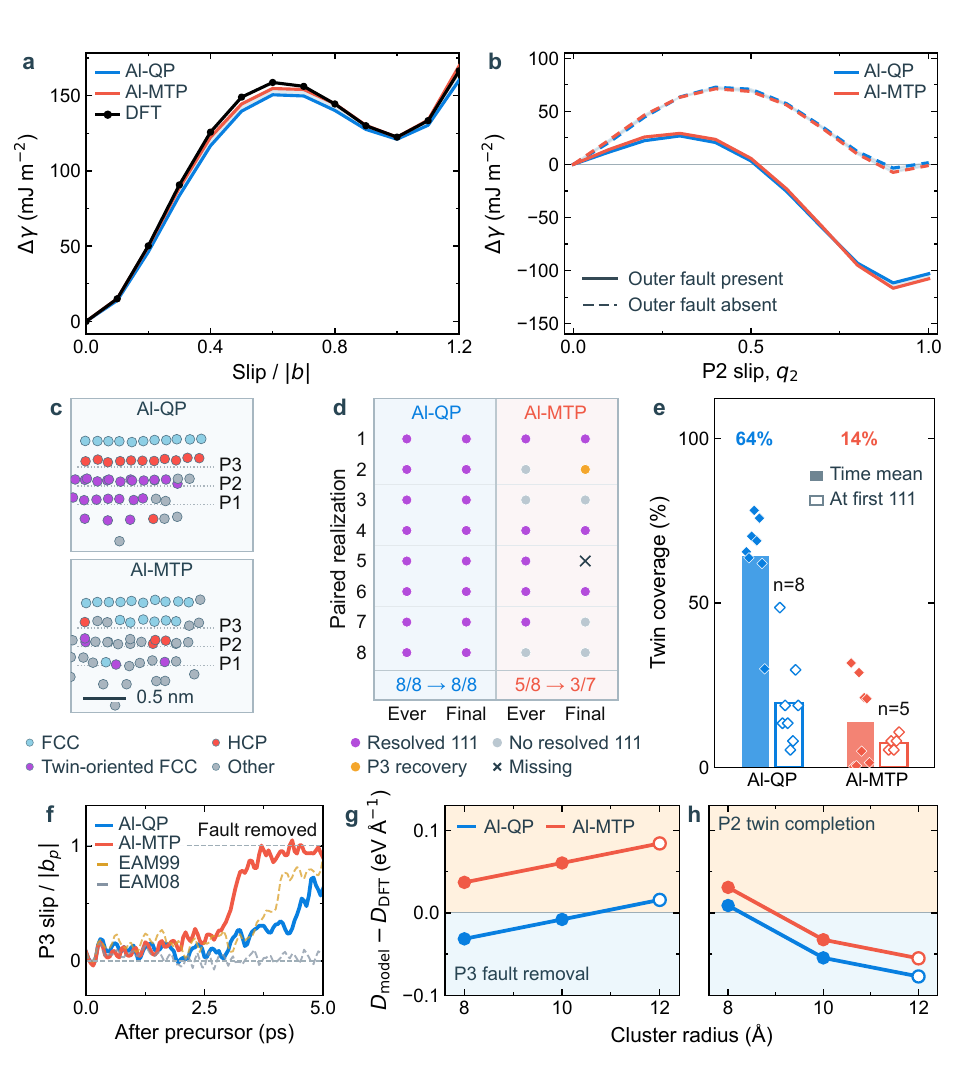}
\par\vspace{5pt}{\fontsize{9}{11}\selectfont\leftskip=0pt\raggedright \textbf{Figure 1 | Matched planar energetics do not determine twin persistence.} \textbf{a}, Zero-load single-plane slip energies for Al-QP and Al-MTP compared with the Andric--Yin--Curtin DFT reference. \textbf{b}, Layered-crystal energy profiles for P2 shear with (solid) and without (dashed) an outer P3 fault. \textbf{c}, Terminal atomic structures at 24.50 ps (realization 2 in \textbf{d}), showing retained central 111 twin stacking under Al-QP (top) and P3 parent recovery under Al-MTP (bottom). \textbf{d}, Occurrence (Ever) and terminal retention (Final) of central 111 stacking across eight paired thermal-velocity realizations. \textbf{e}, Resolved twin coverage from the common precursor: filled bars show time means; open bars, coverage at first central 111 completion; diamonds, individual realizations. \textbf{f}, P3 displacement histories for four potentials from the same precursor. \textbf{g,h}, Model-minus-DFT slip-force imbalance at the P3 fault-removing (\textbf{g}) and P2 twin-completing (\textbf{h}) cores for cluster radii of 8, 10 and 12 Å. Open symbols at 12 Å denote calculations capped at the fixed electronic-step limit (Methods).\par}
\end{figure}

\clearpage

% NEW_FIGURE Fig2
\begin{figure}[p]\centering
\includegraphics[width=\textwidth,height=0.79\textheight,keepaspectratio]{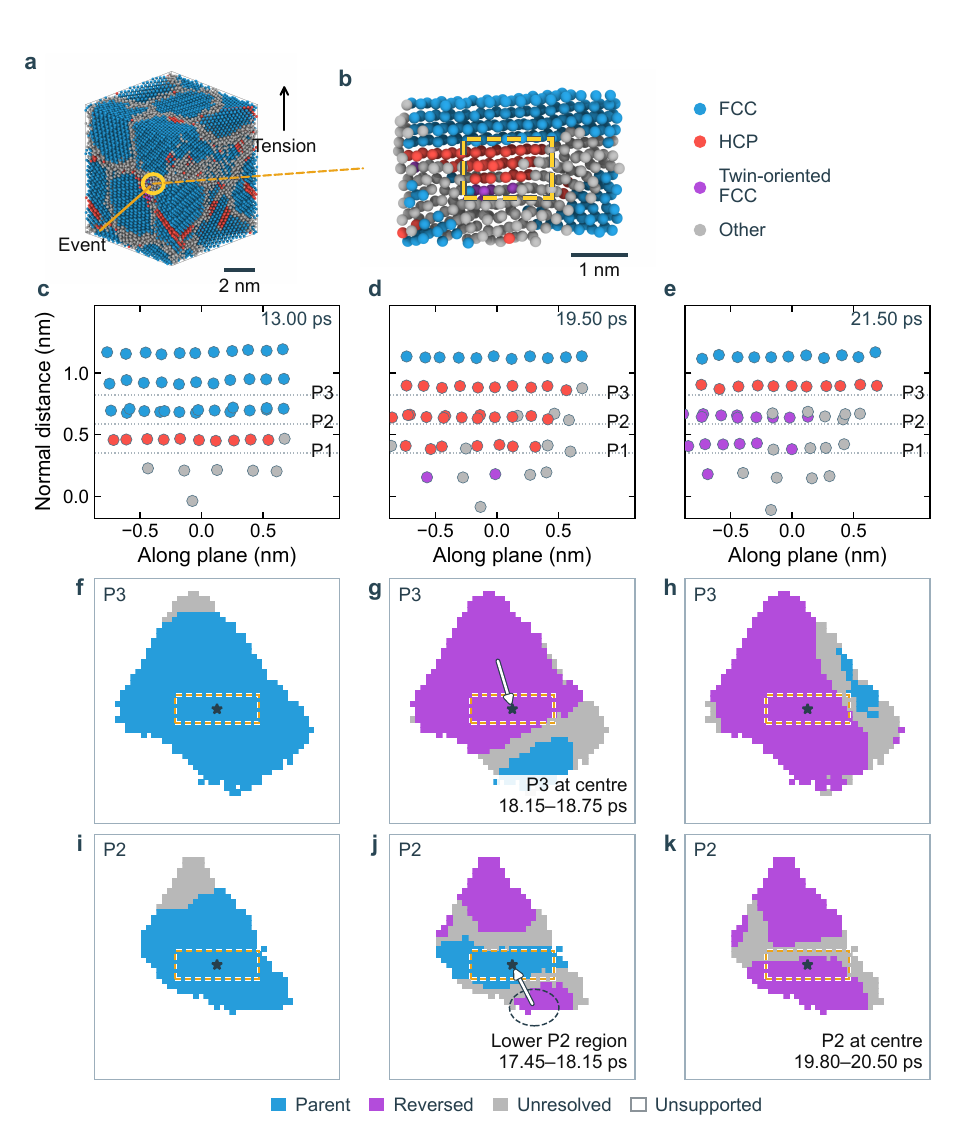}
\par\vspace{5pt}{\fontsize{9}{11}\selectfont\leftskip=0pt\raggedright \textbf{Figure 2 | A separated-fault precursor assembled by asynchronous slip fronts.} \textbf{a}, The 50,846-atom Al polycrystal at 19.50 ps under axial tension; the yellow circle marks the event neighbourhood. \textbf{b}, A 0.64-nm-thick section through that neighbourhood exposing the separated faults and adjacent twin. \textbf{c--e}, Layer sections showing a single fault at 13.00 ps (\textbf{c}), separated faults at 19.50 ps (\textbf{d}) and the completed local twin at 21.50 ps (\textbf{e}), with P1--P3 marking successive slip interfaces. \textbf{f--k}, Corresponding in-plane stacking maps (4.7 nm width) on P3 (\textbf{f--h}) and P2 (\textbf{i--k}) at 13.00, 19.50 and 21.50 ps. Stars mark the central site, yellow dashed boxes the section footprint in \textbf{c--e} and arrows the advancing partial fronts.\par}
\end{figure}

\clearpage

% NEW_FIGURE Fig3
\begin{figure}[p]\centering
\includegraphics[width=\textwidth,height=0.79\textheight,keepaspectratio]{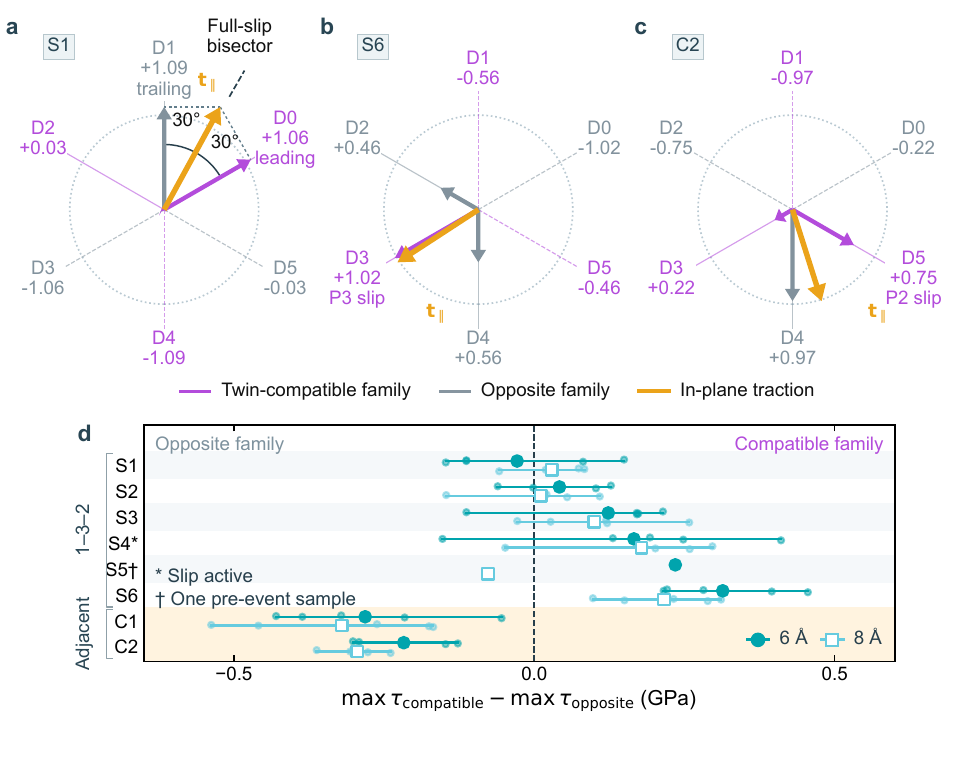}
\par\vspace{5pt}{\fontsize{9}{11}\selectfont\leftskip=0pt\raggedright \textbf{Figure 3 | Directional traction selects the crystallographic family.} \textbf{a--c}, Mean pre-event P3 in-plane traction (gold arrows) and its projections onto the six signed Shockley directions (6 Å averaging radius) for the representative 1--3--2 event S1 (\textbf{a}), a second 1--3--2 site S6 (\textbf{b}) and adjacent twin growth C2 (\textbf{c}). Purple arrows lead from parent to reversed stacking; grey arrows denote the opposite family; dotted circles mark 1 GPa. In \textbf{a}, two 30$^{\circ}$ arcs and the dashed line mark the D0--D1 full-slip bisector (1.3$^{\circ}$ from the mean traction). \textbf{d}, P3 directional preference across six 1--3--2 sites (S1--S6) and two adjacent-plane sites (C1, C2; beige shading), defined as maximum compatible minus maximum opposite-family shear. Large symbols, horizontal bars and small points denote temporal means, full ranges and individual samples, respectively, for 6 Å (circles) and 8 Å (squares) probe radii. An asterisk marks pre-existing slip activity; a dagger marks a single pre-event snapshot.\par}
\end{figure}

\clearpage

% NEW_FIGURE Fig4
\begin{figure}[p]\centering
\includegraphics[width=\textwidth,height=0.79\textheight,keepaspectratio]{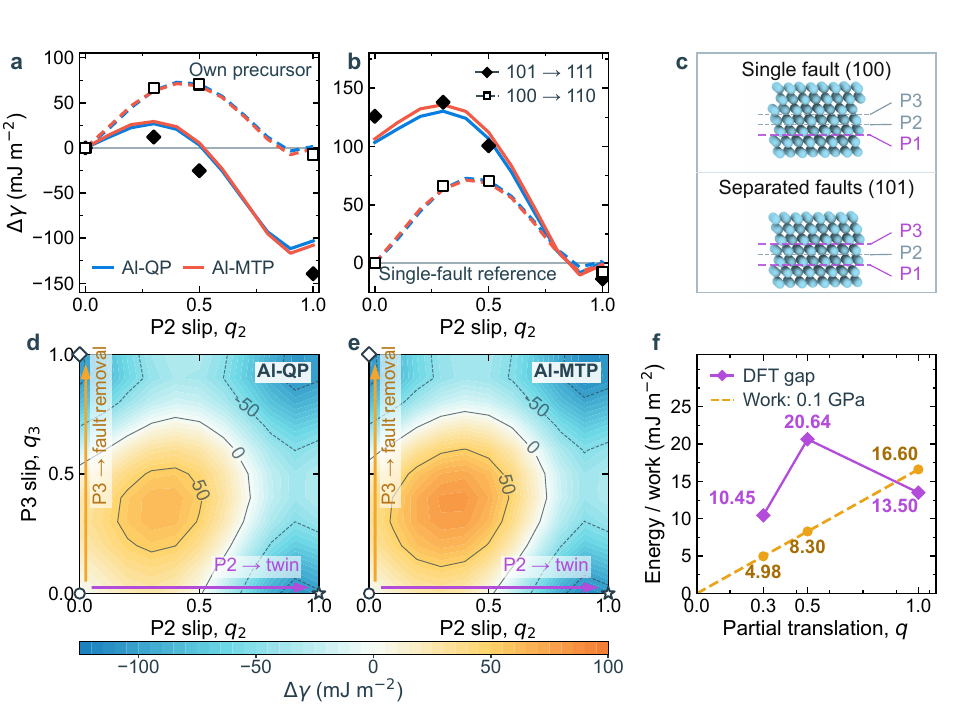}
\par\vspace{5pt}{\fontsize{9}{11}\selectfont\leftskip=0pt\raggedright \textbf{Figure 4 | A narrow energetic margin between twin completion and fault removal.} \textbf{a}, Energy profiles along P2 shear with the outer P3 fault present (solid curves and filled diamonds; 101$\rightarrow$111) or absent (dashed curves and open squares; 100$\rightarrow$110), each referenced to its own relaxed precursor. Symbols denote independently relaxed DFT states. \textbf{b}, The same P2 slip paths referenced to a common single-fault (100) energy, where the 101$\rightarrow$111 series retains the energetic cost of outer-fault preparation. \textbf{c}, DFT-relaxed single-fault (100) and separated-fault (101) reference structures. \textbf{d,e}, Two-dimensional energy surfaces for Al-QP (\textbf{d}) and Al-MTP (\textbf{e}) starting from the 101 precursor (circle) toward twin completion (star) or P3 fault removal (diamond), interpolated over an 11$\times$11 grid. \textbf{f}, DFT energy separation between the fault-removal and twin-completion channels at matched partial translations $q$, compared with the mechanical work of a 0.1 GPa shear stress difference.\par}
\end{figure}

\clearpage

% NEW_FIGURE Fig5
\begin{figure}[p]\centering
\includegraphics[width=\textwidth,height=0.79\textheight,keepaspectratio]{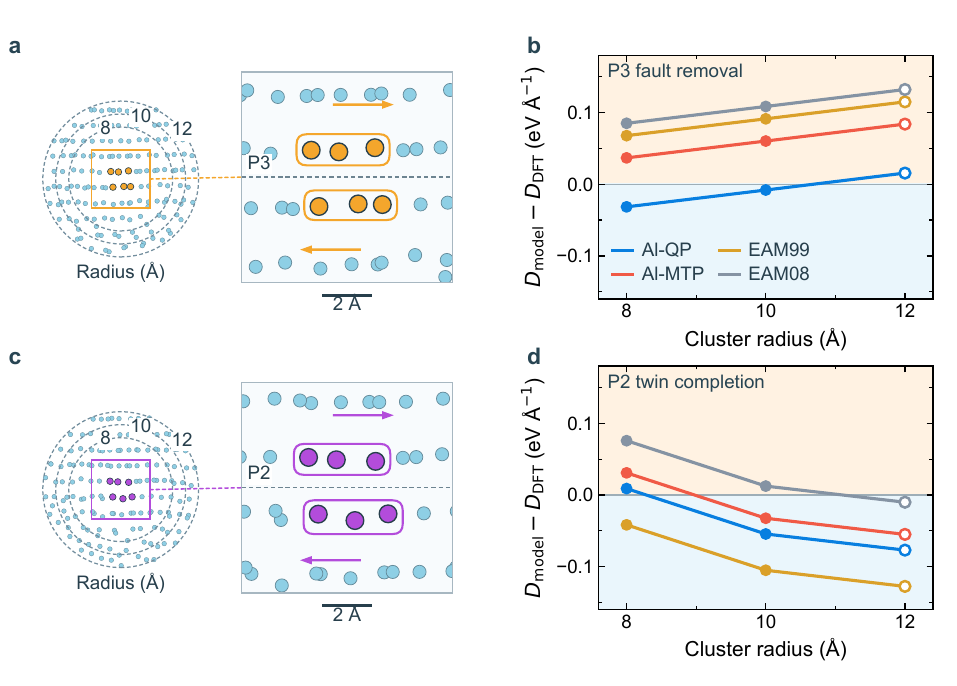}
\par\vspace{5pt}{\fontsize{9}{11}\selectfont\leftskip=0pt\raggedright \textbf{Figure 5 | Forces within moving partial cores carry the divergence between potentials.} \textbf{a,c}, Atomic sections (left; dashed circles at cluster radii of 8, 10 and 12 Å) and enlarged six-atom core views (right) for advancing P3 fault-removing (\textbf{a}) and P2 twin-completing (\textbf{c}) partials. Coloured outlines enclose three core atoms per side; arrows indicate positive relative slip. \textbf{b,d}, Full-system model minus finite-cluster DFT slip-force imbalance at the P3 (\textbf{b}) and P2 (\textbf{d}) cores. Positive values indicate a stronger forward drive than DFT. Open symbols at 12 Å denote calculations capped at the fixed electronic-step limit (Methods).\par}
\end{figure}

\clearpage

% NEW_FIGURE Fig6
\begin{figure}[p]\centering
\includegraphics[width=\textwidth,height=0.79\textheight,keepaspectratio]{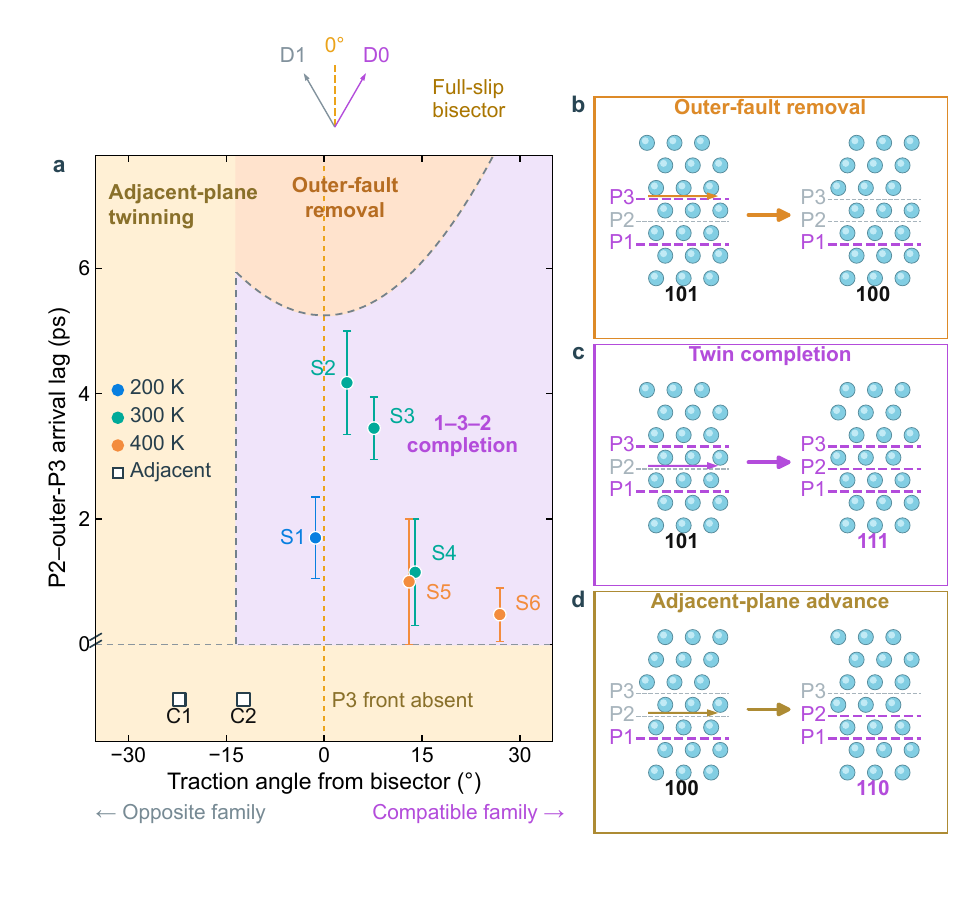}
\par\vspace{5pt}{\fontsize{9}{11}\selectfont\leftskip=0pt\raggedright \textbf{Figure 6 | A directional-kinetic criterion for twin persistence.} \textbf{a}, Regime map in the plane of P3 traction angle (relative to the full-slip bisector) and P2-minus-P3 front-arrival lag. Filled circles mark the six 1--3--2 sites across three polycrystals (coloured by temperature); open squares in the lower band mark adjacent-plane cases where a P3 front is absent; whiskers span temporal sampling intervals. Dashed curves indicate schematic boundaries separating 1--3--2 completion, outer-fault removal and adjacent-plane twinning. \textbf{b--d}, Idealized close-packed layer translations illustrating outer-fault removal (\textbf{b}), twin completion (\textbf{c}) and adjacent-plane advance (\textbf{d}).\par}
\end{figure}

\clearpage

\end{document}